\documentclass[useAMS,usenatbib]{mn2e}
\usepackage{epsfig} 
\def\mgii{Mg~{\sc ii}~} 
\def\mgiia{Mg~{\sc ii}$\lambda$2796~} 
\def\mgiib{Mg~{\sc ii}$\lambda$2803~} 
\def\mgiiab{Mg~{\sc ii}$\lambda\lambda$2796,2803~} 
\def\feii{Fe~{\sc ii}~} 

\def\oiiiab{O~{\sc iii}$\lambda\lambda$4959,5007~} 
\def\oiiia{O~{\sc iii}$\lambda$4959~} 
\def\oiiib{O~{\sc iii}$\lambda$5007~} 
 
\def\civ{C~{\sc iv}~}

\def\halpha{H$\alpha$~}
\def\hbeta{H$\beta$~}

\def\chisq{$\chi^{2}~$} 
\def\kms{km~s$^{-1}~$} 
\def\aj{{AJ}}%
\def\araa{{ARA\&A}}%
\def\apj{{ApJ}}%
\def\apjs{{ApJS}}%
\def\aap{{A\&A}}%
\def\mnras{{MNRAS}}%
\def\pasp{{PASP}}%
\title[Spectra of microvariable
radio-quiet QSOs]{Probing spectral properties of radio-quiet quasars searched for
optical microvariability}
\author[Chand, Wiita \& Gupta]{Hum Chand$^{1}$\thanks{E-mail: hum@aries.res.in (HC); wiita@chara.gsu.edu (PJW);  acgupta30@gmail.com \& alok@aries.res.in (ACG)}, 
Paul J.\ Wiita$^{2,3}$$^{\star}$ and  Alok C.\ Gupta$^{1}$$^{\star}$\\
$^{1}$Aryabhatta Research Institute of Observational Sciences (ARIES),
Manora Peak, Nainital $-$ 263129, India\\
$^{2}$Department of Physics and Astronomy, Georgia State University, Atlanta,
GA 30302--4106, USA\\
$^{3}$School of Natural Sciences, Institute for Advanced Study,
Princeton, NJ 08540, USA}
\begin{document}
\date{Accepted 2009 October 25. Received 2009 October 2; in original form 2009 August 16}

\pagerange{\pageref{firstpage}--\pageref{lastpage}} \pubyear{2010}

\maketitle

\label{firstpage}
\begin{abstract}
We obtained SDSS spectra for a set of 37 radio-quiet quasars (RQQSOs)
that had been previously examined for rapid small scale optical
variations, or microvariability.  Their \hbeta and \mgii emission lines were carefully fit to
determine line widths (FWHM) as well as equivalent widths (EW) due to the 
broad emission line components.  The line widths were used to
estimate black hole masses and Eddington ratios, $\ell$.  Both EW and
FWHM are anticorrelated with $\ell$.  The EW distributions provide no
evidence for the hypothesis that a weak jet component in the RQQSOs is
responsible for their microvariability.
\end{abstract}
\begin{keywords}
galaxies: active -- quasars: emission lines -- quasars: general
\end{keywords}
\section{Introduction}
Over the past 15 years there have been rather extensive examinations
of a significant sample of radio-quiet QSOs (RQQSOs) and Seyfert
galaxies for small brightness changes (typically 0.02 mag) over short
times (a few hours) (e.g., Stalin et al.\ 2004b; Carini et al.\ 2007;
Ram{\'i}rez et al.\ 2009). This phenomenon of microvariability, or
intranight optical variability (INOV) was first confirmed for blazars
(e.g., Miller, Carini \& Goodrich 1989; Carini 1990) for which
microvariability almost certainly arises from the relativistic jet
(e.g., Marsher, Gear \& Travis 1992), at least when the source is in an 
active state; however, in low states it is possible that rapid fluctuations are due to processes originating in or just above
the accretion disc (for a review, see Wiita 2006).  Since RQQSOs lack
significant jets, the microvariability in radio-quiet objects may arise
from processes on the accretion disc itself, and thus could possibly
be used to probe the discs (e.g., Gopal-Krishna, Wiita \&
Altieri 1993; Stalin et al.\ 2004a; Kelly, Bechtold \& Siemiginowska 2009).

Recently Carini et al.\ (2007) reported  new observations for several
sources and also  compiled a sample of 117 objects
from the literature which have been searched for microvariability
(their Table 3). Of these, 47 are classified as Seyfert galaxies, 6 as
broad absorption line (BAL) QSOs, and 64 as QSOs. In addition, in order to learn which classes
might be the most fruitful for making further searchs for microvariability they
also classified their sample in terms of following: redshift distribution; radio
loudness, through $R$, the ratio of the radio [5 GHz] flux to the
optical [4400\AA] flux); optical magnitude, $m_V$; luminosity; and observing strategy. In their entire sample 21.4\% of the objects were found to
exhibit microvariability, but among objects classified as Seyfert
galaxies, BAL QSOs and QSOs, microvariability was found in 17\%, 50\%
and 23.2\%, respectively (Carini et al.\ 2007).  The observed high
fraction of microvariations in BAL QSOs (although the sample is quite
small) suggests that while planning microvariability studies to
investigate the physical processes in or near the accretion disc, it
might be worthwhile to invest more observing time on the BALQSO class. \par

  With improvements in observation quality as well as in sample size,
  constraints on the models capable of producing the intranight variability have
  also improved over last decade.  Recently, Czerny et al.\ (2008)
  have used non-simultaneous optical and X-ray data of  10 RQQSOs with confirmed INOV to compare observational
  constraints on the variability properties with the predictions of
  theoretical models such as: (i) irradiation of an accretion disc by
  a variable X-ray flux (e.g., Rokaki, Collin-Souffrin \& Magnan 1993;
  Gaskell 2006); (ii) an accretion disc instability (e.g., Mangalam \&
  Wiita 1993); (iii) the presence of a weak Doppler boosted jet, or
  ``blazar component'' (e.g., Gopal-Krishna et al.\ 2003). Their
  investigation suggests that a blazar component model yields
  the highest probability of detecting INOV.

In this blazar component scenario, spectral properties of the sources
can play a crucial role in constraining the models further. For
instance, if blazar components are dominating the variability of
RQQSOs, then, due to the increase in the continuum level, one would
expect emission lines to be diluted.  Therefore smaller equivalent
widths (EWs) of prominent emission lines such as \hbeta and \mgii
should be detected in sources that showed microvariability when
compared to their average values in the whole sample. If we take the
extreme case of BL Lacertae objects, which often lack observable
emission lines and are usually defined as objects that have no
emission line with an EW $\geq$5\AA \ (e.g., Stickel et al.\ 1991;
cf., March{\~a} et al.\ 1996), this
dilution by the jet component is understood to be severe.  So it
becomes very important to test whether microvariability of RQQSOs has
any correlation with spectral parameters such as EW and full width at
half maximum (FWHM) of prominent emission lines. The \hbeta and \mgii
emission lines are very promising for such investigations, as these
lines have also been found to be very useful in estimating other key
parameters of AGN central engines such as black hole (BH) mass (e.g.,
McGill et al.\ 2008) and Eddington ratio (e.g., Dong et al.\ 2009a,
2009b). The average \hbeta EW of a large sample of quasars is found to
be around 62.4\AA \ (Foster et al.\ 2001), so any correlation of \hbeta
EW or FWHM {\bf would} not only give insight about the nature of variability,
such as the presence or absence of blazar components, but also {\bf would} be
very useful for making a promising sample for future microvariability
studies.  Similarly the measurements of BH masses for RQQSOs that show
microvariability may well be another important constraint {\bf on} the
models trying to understand the nature of their optical
microvariability.
 
Here we have worked toward these goals by exploiting the optical
spectra available from Sloan Digital Sky Survey (SDSS) Data Release 7
(DR7; Abazajian et al.\ 2009) with careful spectral modeling of the
\hbeta and \mgii emission line regions. First we aim to investigate
any effect of these key spectral parameters (e.g., EW and FWHM) on
microvariability of RQQSOs. Second we estimate other relevant
AGN parameters such as the black hole mass, M$_{bh}$, and the
Eddington ratio in the context of our RQQSOs sample, which has been
extensively searched for microvariability. \par

The paper is organized as follows. Section 2 describes the data sample
and selection criteria while Section 3 gives details of our spectral
fitting procedure. In Section 4 we focus on BH mass measurements and
in Section 5 we give estimates of Eddington ratios and of BH growth
times.  Section 6 gives a discussion and conclusions.  Throughout, we
have used a cosmology with $H_{\rm 0}$=70
km\,s$^{-1}$\,Mpc$^{-1}$, $\Omega_{\rm M}$=0.3 and $\Omega_{\rm
  \Lambda}$=0.7. 

\begin{table*}
 \centering
 \begin{minipage}{140mm}
\caption{ Our sample of radio-quiet QSOs and Seyfert
galaxies from the compilation of Carini et al.\ (2007)}
\label{lab:tabsamp}
\begin{tabular}{@{}llllllllllcl@{}} 
\hline 
\multicolumn{1}{c}{QSO Name\footnote{Name from V{\'e}ron-Cetty \& V{\'e}ron (2006)}}   & {$z_{em}$}   
& \multicolumn{3}{c}{$\alpha_{2000}$} 
& \multicolumn{3}{c}{$\delta_{2000}$} 
&{m$_{V}$}
& {M$_{V}$} 
&  {Variable?} 
& {Class} \\
\hline 
\\

Mrk 1014            & 0.163        &   01& 59& 50.2 &   $+$00& 23& 41&   15.9 &   $-$23.8 &   Y&  Sy 1    \\
US 3150              & 0.467        &   02& 46& 51.9 &   $-$00& 59& 31&   17.1 &   $-$25.0 &  N&  NLS1  \\
US 3472              & 0.532        &   02& 59& 37.5 &   $+$00& 37& 36&   16.8 &   $-$25.7 &  N&  QSO     \\
Q  J0751$+$2919      & 0.912        &   07& 51& 12.3 &   $+$29& 19& 38&   16.2 &   $-$27.7 &  Y&  QSO     \\
PG 0832$+$251        & 0.331        &   08& 35& 35.9 &   $+$24& 59& 41&   16.1 &   $-$25.5 &  Y&  QSO     \\
US 1420              & 1.473        &   08& 39& 35.1 &   $+$44& 08& 11&   17.5 &   $-$27.3 &  N&  QSO     \\
US 1443              & 1.564        &   08& 40& 30.0 &   $+$46& 51& 13&   17.2 &   $-$27.8 &  N&  QSO     \\
US 1498              & 1.406        &   08& 42& 15.2 &   $+$45& 25& 44&   17.7 &   $-$26.7 &  N&  QSO     \\
US 1867              & 0.513        &   08& 53& 34.2 &   $+$43& 49& 01&   16.9 &   $-$25.2 &  N&  Sy 1    \\
PG 0923$+$201        & 0.192        &   09& 25& 54.7 &   $+$19& 54& 04&   15.5 &   $-$24.8 &  Y&  Sy 1    \\
PG 0931$+$437        & 0.456        &   09& 35& 02.6 &   $+$43& 31& 11&   16.0 &   $-$26.4 &  N&  QSO     \\
PG 0935$+$416        & 1.966        &   09& 38& 57.0 &   $+$41& 28& 21&   16.8 &   $-$28.8 &  N&  QSO     \\
CSO 233              & 2.030        &   09& 39& 35.1 &   $+$36& 40& 01&   18.4 &   $-$27.3 &  N&  QSO     \\
CSO 18               & 1.300        &   09& 46& 36.9 &   $+$32& 39& 51&   17.0 &   $-$27.9 &  N&  QSO     \\
US 995               & 0.226        &   09& 48& 59.4 &   $+$43& 35& 18&   16.9 &   $-$23.4 &  Y&  QSO     \\
PG 0946$+$301        & 1.220        &   09& 49& 41.1 &   $+$29& 55& 19&   16.2 &   $-$28.2 &  N&  BAL QSO \\
CSO 21               & 1.190        &   09& 50& 45.7 &   $+$30& 25& 19&   17.3 &   $-$27.3 &  N&  QSO     \\
TON 34               & 1.925        &   10& 19& 56.6 &   $+$27& 44& 02&   15.7 &   $-$29.8 &  Y&  QSO     \\
PG 1049$-$005        & 0.357        &   10& 51& 51.5 &   $-$00& 51& 17&   15.8 &   $-$25.7 &  N&  Sy 1    \\
TON 52               & 0.434        &   11& 04& 07.0 &   $+$31& 41& 11&   17.3 &   $-$24.9 &  Y&  QSO     \\
PG 1206$+$459        & 1.155        &   12& 08& 58.0 &   $+$45& 40& 36&   15.7 &   $-$28.4 &  N&  QSO     \\
UM 497               & 2.022        &   12& 25& 18.4 &   $+$02& 06& 57&   17.7 &   $-$28.0 &  N&  QSO     \\
PG 1248$+$401        & 1.032        &   12& 50& 48.3 &   $+$39& 51& 40&   16.3 &   $-$27.6 &  N&  QSO     \\
CSO 174              & 0.653        &   12& 51& 00.3 &   $+$30& 25& 42&   17.0 &   $-$26.1 &  N&  QSO     \\
CSO 179              & 0.782        &   12& 53& 17.5 &   $+$31& 05& 50&   17.0 &   $-$26.4 &  N&  QSO     \\
Q 1252$+$0200        & 0.345        &   12& 55& 19.7 &   $+$01& 44& 12&   16.2 &   $-$25.3 &  Y&  QSO     \\
PG 1254$+$047        & 1.018        &   12& 56& 59.9 &   $+$04& 27& 34&   16.3 &   $-$27.6 &  N&  BAL QSO \\
PG 1259$+$593        & 0.472        &   13& 01& 12.9 &   $+$59& 02& 07&   15.9 &   $-$26.2 &  N&  QSO     \\
PG 1307$+$085        & 0.154        &   13& 09& 47.0 &   $+$08& 19& 49&   15.1 &   $-$24.6 &  N&  Sy 1    \\
PG 1309$+$355        & 0.183        &   13& 12& 17.7 &   $+$35& 15& 20&   15.6 &   $-$24.6 &  N&  Sy 1.2  \\
CSO 879              & 0.549        &   13& 21& 15.8 &   $+$28& 47& 20&   16.7 &   $-$25.8 &  N&  QSO     \\
PG 1338$+$416        & 1.204        &   13& 41& 00.8 &   $+$41& 23& 14&   16.8 &   $-$27.5 &  N&  QSO     \\
CSO 448              & 0.316        &   14& 24& 55.6 &   $+$42& 14& 05&   17.0 &   $-$24.3 &  Y&  QSO     \\
PG 1444$+$407        & 0.267        &   14& 46& 46.0 &   $+$40& 35& 06&   16.1 &   $-$24.7 &  N&  Sy 1    \\
PG 1522$+$101        & 1.328        &   15& 24& 24.5 &   $+$09& 58& 30&   16.2 &   $-$28.4 &  N&  QSO     \\
Q 1628.5$+$3808      & 1.461        &   16& 30& 13.6 &   $+$37& 58& 21&   17.7 &   $-$27.2 &  Y&  QSO     \\   
PG 1630$+$377        & 1.478        &   16& 32& 01.1 &   $+$37& 37& 49&   16.3 &   $-$28.6 &  N&  QSO     \\
\hline
\end{tabular}
\end{minipage}
\end{table*}  
\section{Data sample and selection criteria}

  Our sample is chosen from the compilation of 117 radio-quiet AGNs in
  Carini et al.\ (2007). These sources have all been searched, often
  extensively, for optical microvariability.  When there have been
  several attempts at monitoring the same RQQSO or Seyfert galaxy by
  different groups of authors around the globe they often have used
  different popular names or even different IAU names for the same
  source. Therefore we first looked in all the papers cited in the
  compiled data (Carini et al.\ 2007, their Table 3) and obtained the
  right ascensions ($\alpha_{2000.0}$) and declinations($\delta_{2000.0}$) of all the sources monitored by
  different groups involved in  searches for their optical microvariability.
   Then we searched these sources in the recent
  catalog of AGN and Quasars by V{\'e}ron-Cetty \& V{\'e}ron (2006) and made
  the list of all sources monitored until 2005 by giving their popular
  name from V{\'e}ron-Cetty \& V{\'e}ron (2006).

We searched for consistently produced optical spectra of these sources 
by looking for these targets in SDSS DR7
(Abazajian et al.\ 2009) and found that out of these 117 sources (Carini
et al.\ 2007) SDSS DR7 spectra were available for the 37 sources listed in
Table~\ref{lab:tabsamp}.  The first four columns give identification information and redshift, the next two provide apparent and absolute
V magnitudes and the last two indicate whether microvariability was detected and whether the radio quiet AGN has been classified as
a QSO, BALQSO, Seyfert (Sy) galaxy,
or Narrow Line Seyfert 1 (NLS1) galaxy. From among the 37 sources in our sample, 8
source spectra cover both \mgii and \hbeta lines, 11 spectra covered
only \hbeta lines while 18 spectra covered only \mgii doublet lines.
This set of RQQSOs becomes an unique sample to look for any dependence
on microvariability of their optical spectral properties as its
members had been already targeted to search for optical
microvariability in the past.
\begin{figure*}
 \epsfig{figure=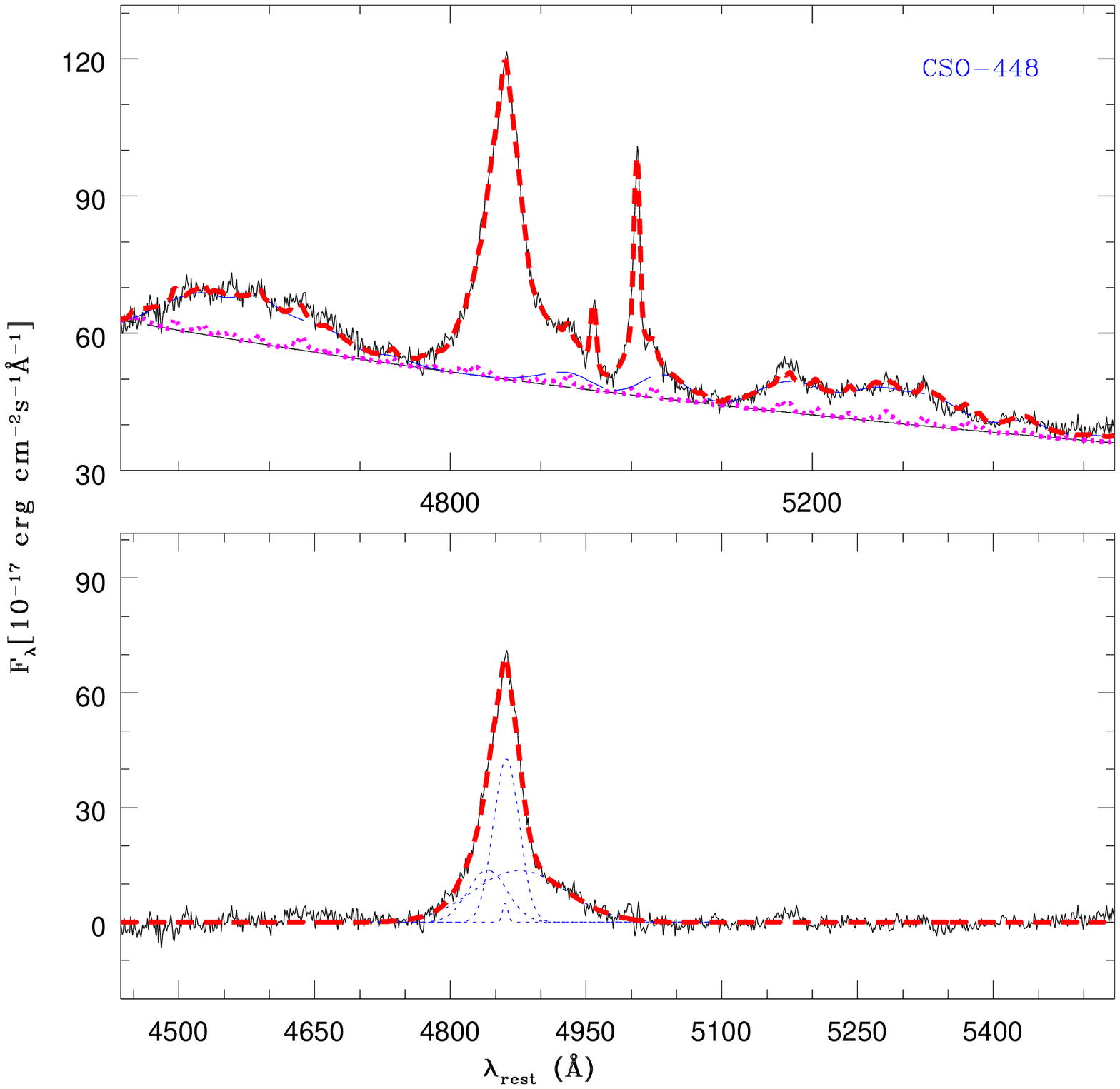,height=12.cm,width=16.cm,angle=0} 
 \caption{The best fit to the \hbeta emission line of the 
SDSS spectra of the QSO CSO 448.  Upper panel:
complete spectrum fit (thick dashed/red) and components of the fit:
power law continuum (thin dashed/black line), broad \feii\ (dot-dashed/blue),
narrow \feii\ (dotted/magenta)  lines. Lower panel:
continuum, \feii\ and metal line subtracted spectrum (solid/black)
  with the best fit total \hbeta profile (thick dashed/red) and
  \hbeta\ components (dotted/blue) lines.  Note that the entire fit is
  performed simultaneously (not first continuum subtraction then
  \hbeta fit) but these aspects are shown separately for the sake of clarity.}
\label{lab:fig_hbetademo}
\end{figure*}
\begin{figure*}
 \epsfig{figure=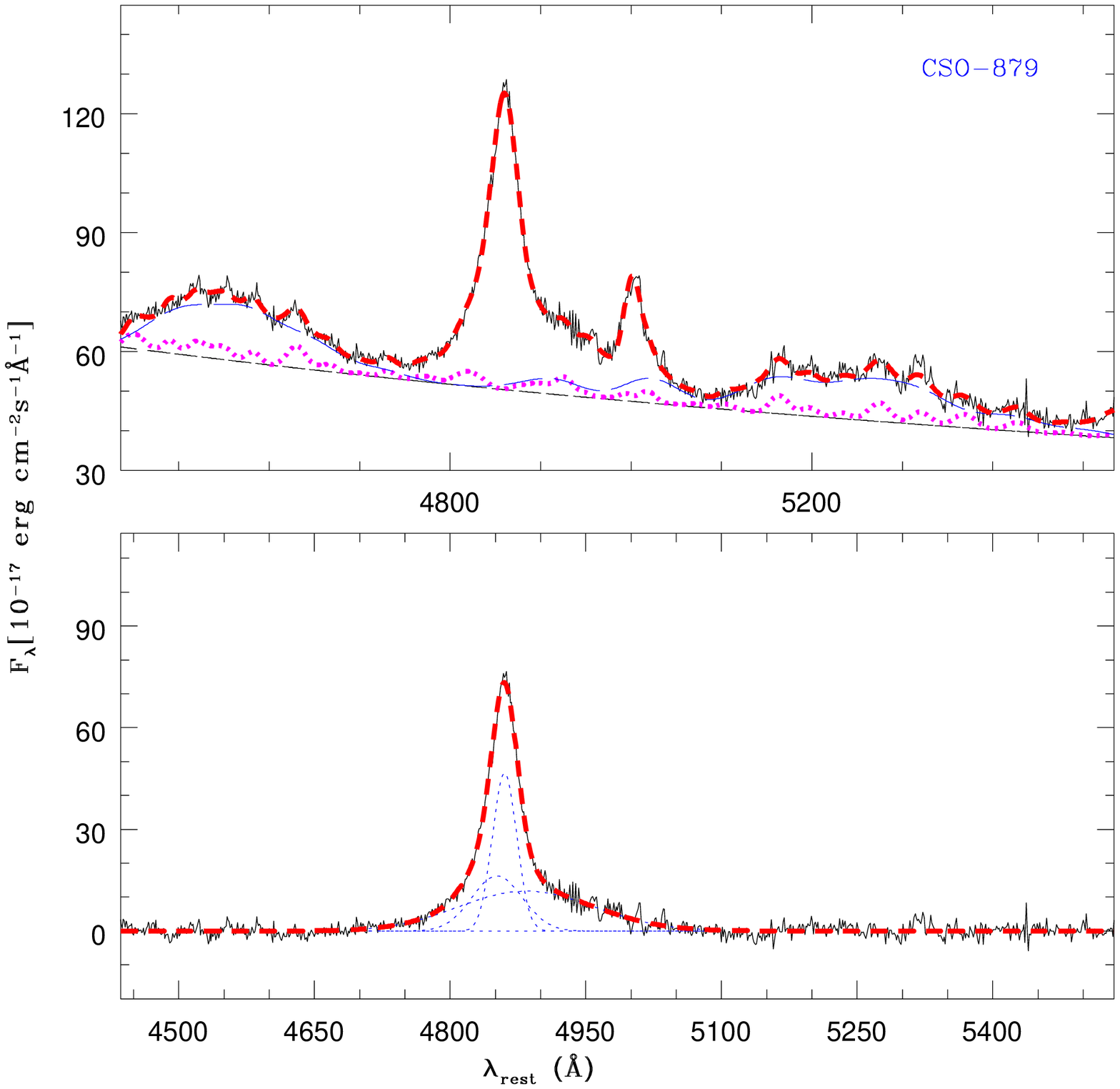,height=12.cm,width=16.cm,angle=0} 
 \caption{ Same as  Fig.~\ref{lab:fig_hbetademo} for CSO 879.}
\label{lab:fig_hbetademo2}
\end{figure*}
\section{Line analysis through simultaneous spectral fitting}
\label{sec:aalysis}
The spectra were first corrected for Galactic extinction using the
extinction map of Schlegel et al.\ (1998) and the reddening curve of
Fitzpatrick (1999). Then they were transformed into the
rest frame using the redshift given in the header of the SDSS spectra.\par
 
 In order to have accurate measurements of spectral properties such as the EWs and
FWHMs of emission lines, one has to carefully take into account the
contribution from continuum as well as from \feii emission
multiplets.  The common practice was to first  fit and
subtract the AGN continuum and then the \feii emission (e.g., Boroson \&
Green 1992; Marziani et al.\ 2003) and finally the other emission lines.
The other possibility is to fit simultaneously the 
continuum and the \feii emission multiplets and other emission 
lines (e.g., Dong et al.\ 2008, 2009a).
Unfortunately, for the optical spectra of most of RQQSOs in our sample,
fitting the continuum and the \feii emission is very complicated because: (i) there are essentially no emission-line free regions
where the continuum can be determined (Vanden Berk et al.\ 2001); (ii)
the prominence of \feii features and their blending with the \hbeta
and \mgii lines; (iii) the \hbeta line is highly blended with the \oiiiab
lines.   
As a result, we have opted to carry out simultaneous
fits\footnote{To carry out the simultaneous fit we have used the \textsc{MPFIT}
  package for nonlinear fitting, written in \textsc{Interactive Data Language}
  routines.  MPFIT is kindly provided by Craig B. Markwardt and is
  available at http://cow.physics.wisc.edu/\~{}craigm/idl/.} of
continuum, \feii emission, \hbeta and \mgii and all other metal emission lines 
present in the spectra.
 
 For this purpose we follow the following procedure.
\subsection{\hbeta\ region fit} 
\label{subsec:hbeta}
For fitting the optical region comprising the \hbeta line, we
consider the rest wavelength range between 4435 and 5535~\AA.  The
continuum is modeled by a single power law, as since we limit {\bf ourselves} to
below 5600\AA\, there is no need of a double or broken power law fit
(e.g, Dong et al.\ 2008). As the profile of the \hbeta line is rather complex, it
is fitted with multiple (as many as four) Gaussians. As an initial
guess we consider two narrow, one broad and one very broad
components. To reduce the arbitrariness of the component fits and also
to make the decomposition more physical, we have constrained the
redshift and width of the two narrow components of \hbeta to be the same as
those of the \oiiiab lines.  The line profile of \oiiib (and hence of
\oiiia) are modeled as double Gaussians, with one stronger narrow
component with width less than 2000 \kms\ and one weaker broader
component with width less than 4000 \kms. However, if for some source
spectra the second component is not statistically required, the
procedure we use automatically drops it during the fit. So, basically
we have only four free parameters for the width and redshift of
\oiiiab and narrow \hbeta\ components (two for redshift and two for
width).  \par
\begin{figure*}
\epsfig{figure=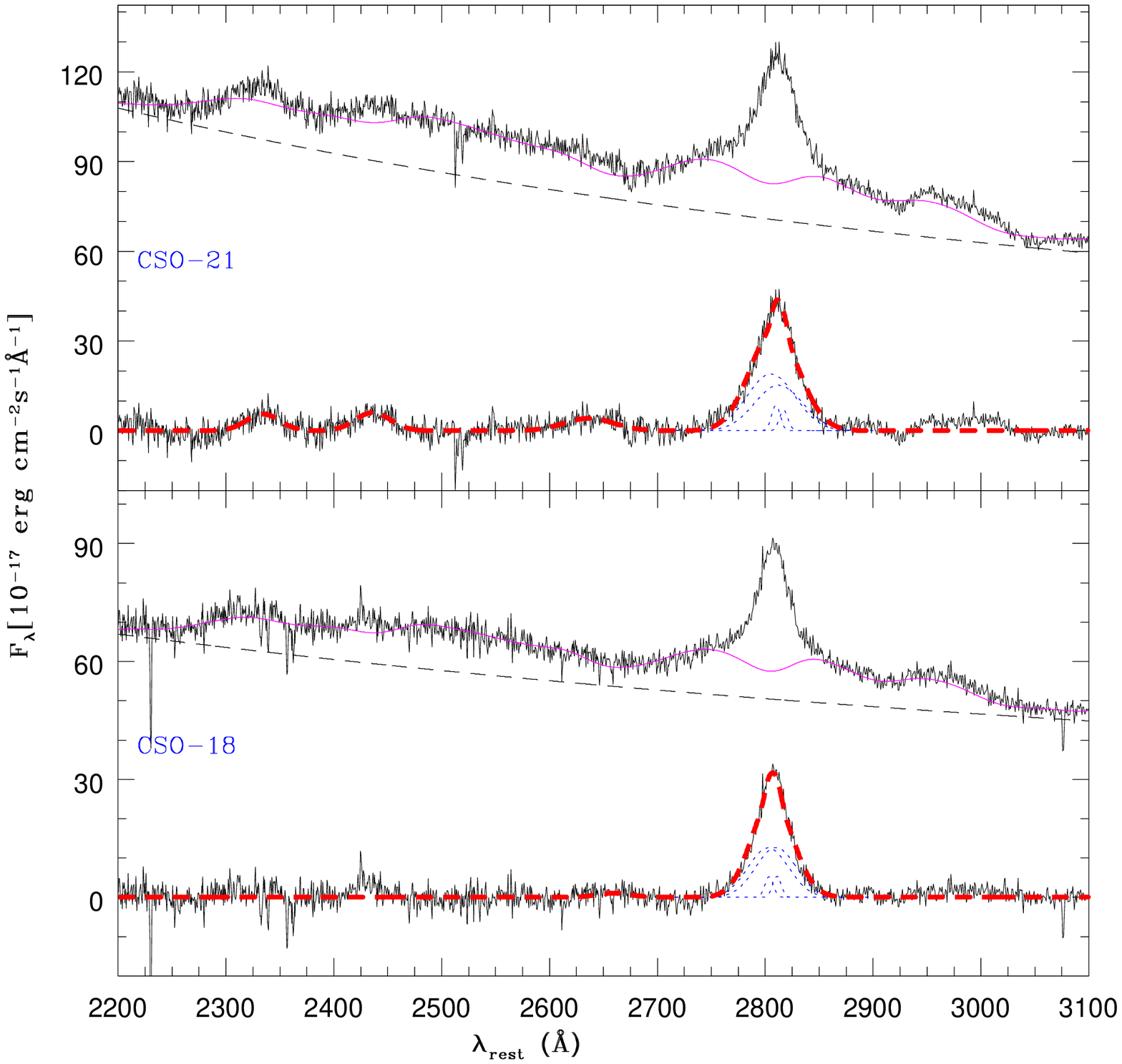,height=12.cm,width=16.cm,angle=0}
\caption[]{\mgii emission line fits for two RQQSOs, CSO 21 and
  CSO 18. The upper plot in each panel shows the SDSS spectra and
  continuum as a dashed line with the best fit UV \feii\ template as a
  solid (magenta) line.  The lower plot in each panel show the
  continuum and \feii\ subtracted spectrum (solid) the best fit
  \mgii profile (thick dashed/red) and \mgiiab\ components by (dotted/blue)
  lines.  Note that the entire fits are performed simultaneously
  (not first continuum subtraction, then \mgii fit) but are shown
  separately for the sake of clarity. }
\label{lab:fig_mgiidemo}
\end{figure*}
\begin{figure*}
 \epsfig{figure=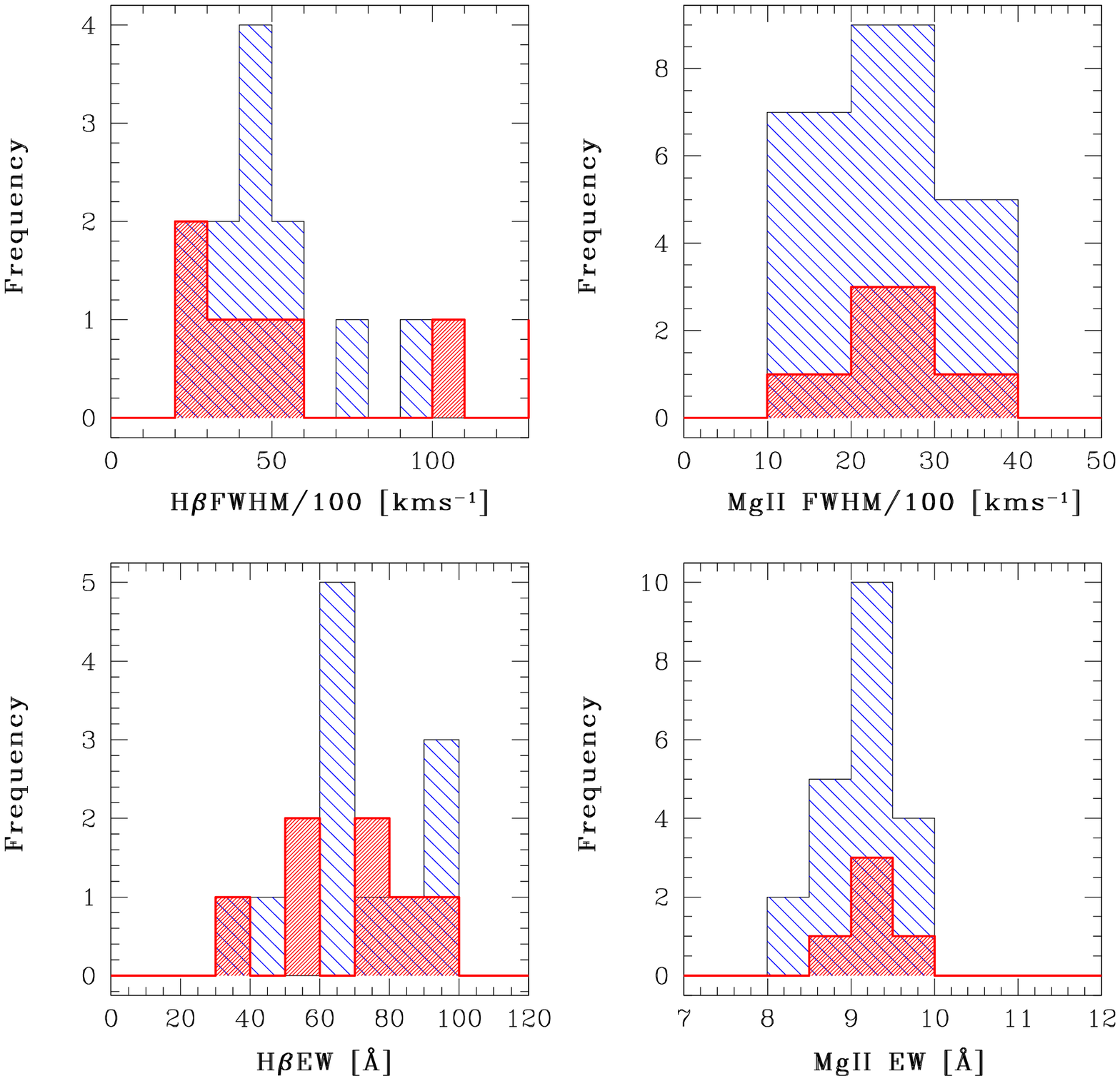,height=12.cm,width=16.cm,angle=0} 
 \caption{Histograms of
FWHM and rest frame equivalent width (EW) based on best fits of
\hbeta and \mgii lines. The shaded regions correspond to sources with
confirmed optical microvariability while the non-shaded regions are for those sources
for which optical microvariability is not detected.}
\label{lab:histo_ewfwhm}
\end{figure*}
  The optical \feii\ emission is modeled with two separate sets of
  templates in analytical forms similar to those used by Dong et al.\ (2008). 
  One template is for the broad-line system and the
  other for the narrow-line system, and we assume the form $C(\lambda) =
  c_{b}C_{b}(\lambda) + c_{n}C_{n}(\lambda)$, where $C_{b}(\lambda)$
  represents the broad \feii lines and $C_{n}(\lambda)$ the narrow \feii
  lines, with the relative intensities fixed at those of I\,ZW\,1, as
  given in Tables A.1 and A.2 of V{\'e}ron-Cetty, Joly \& V{\'e}ron (2004).  The
  redshift of the broad \feii lines and the \hbeta\ component is fitted
  as a free parameter while their widths are kept the same, with a constraint
  that they should be larger than 1000 \kms; this width is used
  for estimating the BH masses.  Similarly for the narrow \feii
  line the redshift is fitted as a free parameter but the width is kept 
the same
  as that of the stronger narrow \oiiib component. The fourth (very broad) \hbeta component,
  if required for the fit, is subject only to the 
  constraint that its width should be more than 1000 \kms. The
  emission lines other than \feii and \oiiiab  (see Table 2 in
  Vanden Berk et al.\ 2001), are modeled with single Gaussians.  The
  final fit is achieved by simultaneously varying all the free
  parameters to minimize the \chisq\ value until the reduced $\chi^2_r$ is
  $\approx 1$.  Samples  of our spectral fitting in the optical
  region are given for \hbeta in Fig.~\ref{lab:fig_hbetademo} and 
  Fig.~\ref{lab:fig_hbetademo2}. The values for FWHM and EW (both
the broad component, EW$_B$, and total, EW$_{all}$) for the \hbeta
lines are given in Table~\ref{lab:tabhbeta}. \par

\begin{table*}
 \centering
 \begin{minipage}{140mm}
  \caption{Results of the \hbeta line analyses}
  \label{lab:tabhbeta}

  \begin{tabular}{@{}lrrrccrr@{}}

\hline
\multicolumn{1}{c}{QSO Name}     & \multicolumn{1}{c}{z } &
                         \multicolumn{1}{c}{$\frac{L(5100\AA)}{10^{44} \rm erg~s^{-1}}$ }    &
                         \multicolumn{1}{c}{FWHM(km s$^{-1}$)}               &
                         \multicolumn{1}{c}{log($\frac{M_{bh}}{M_{\odot}}$)} &
                         \multicolumn{1}{c}{$\frac{L_{bol}}{L_{edd}}$} &
                         \multicolumn{1}{c}{EW$_B$(\AA)} &
                         \multicolumn{1}{c}{EW$_{all}$(\AA)} \\
 \hline
 CSO 174	& 0.653 & 18.41 &  7034.92 & 9.24&0.07 &  10.42 & $ 95.31 \pm 0.98 $ \\
 CSO 179	& 0.782 & 52.71 &  5039.39 & 9.18&0.24 &  47.80 & $ 87.87 \pm 1.21 $ \\
 CSO 448	& 0.317 &  7.45 &  3133.75 & 8.34&0.23 &  14.42 & $ 78.97 \pm 0.66 $ \\
 CSO 879	& 0.551 & 27.24 &  4072.64 & 8.85&0.26 &  21.36 & $ 91.02 \pm 0.67 $ \\
 Mrk 1014  	& 0.163 &  3.77 &  2393.82 & 7.96&0.28 &  30.69 & $ 53.53 \pm 0.51 $ \\
 PG 0832$+$251	& 0.331 & 13.89 &  5059.36 & 8.89&0.12 &  20.72 & $ 84.59 \pm 0.55 $ \\
 PG 0923$+$201	& 0.192 &  8.85 & 10461.15 & 9.42&0.02 &  48.81 & $ 75.77 \pm 0.49 $ \\
PG 0931$+$437	& 0.458 & 31.42 &  9537.59 & 9.62&0.05 &  11.47 & $ 68.42 \pm 0.52 $ \\
 PG 1049$-$005  & 0.359 & 24.17 &  5193.56 & 9.03&0.15 &  22.16 & $ 76.06 \pm 0.40 $ \\
 PG 1259$+$593 	& 0.474 & 42.13 &  2922.88 & 8.65&0.63 &   4.33 & $ 41.06 \pm 0.50 $ \\
 PG 1307$+$085	& 0.154 &  5.22 &  4080.08 & 8.49&0.11 &  62.77 & $ 99.83 \pm 0.54 $ \\
 PG 1309$+$355	& 0.183 & 10.98 &  4326.57 & 8.70&0.15 &  20.86 & $ 35.27 \pm 0.46 $ \\
 PG 1444$+$407 	& 0.268 & 10.15 &  3528.12 & 8.51&0.21 &   8.26 & $ 66.90 \pm 0.50 $ \\
 Q 1252$+$0200 	& 0.344 & 16.62 & 23694.59 &10.27&0.01 &  26.78 & $ 90.09 \pm 0.54 $ \\
 TON 52	  	& 0.434 & 16.68 &  2724.97 & 8.39&0.46 &  11.90 & $ 38.78 \pm 0.69 $ \\
 US 1867	& 0.514 & 25.11 &  2837.29 & 8.52&0.52 &  33.89 & $ 68.80 \pm 0.46 $ \\
 US 3150	& 0.469 & 14.44 &  3402.68 & 8.55&0.27 &  14.78 & $ 68.28 \pm 0.67 $ \\
 US 3472	& 0.532 & 25.48 &  4587.36 & 8.94&0.20 &  14.30 & $ 69.93 \pm 0.64 $ \\
 US 995     	& 0.227 &  3.90 &  4509.77 & 8.51&0.08 &  46.58 & $ 51.95 \pm 0.58 $ \\
\hline
\end{tabular}
\end{minipage}
\end{table*}
\begin{table*}
 \centering
 \begin{minipage}{140mm}
  \caption{Results of the \mgii line analyses}
  \label{lab:tabmgii}

  \begin{tabular}{@{}lrcclccrr@{}}

\hline
\multicolumn{1}{c}{QSO Name}     & \multicolumn{1}{c}{z } &
                          \multicolumn{1}{c}{$\frac{L(3000\AA)}{10^{44} \rm erg~s^{-1}}$ }         &
                         \multicolumn{1}{c}{FWHM(kms$^{-1}$)}              &
                         \multicolumn{1}{c}{Log($\frac{M_{bh}}{M_{\odot}}$)\footnote{Using the McLure \& Dunlop (2004) scaling relation, i.e., Eq.~\ref{logMuv_L30.eq}. The values given in parentheses are from our \hbeta\ measurements for the same object from Table \ref{lab:tabhbeta}.}} &
                         \multicolumn{1}{c}{Log($\frac{M_{bh}}{M_{\odot}}$)\footnote{Using the fixed slope of the $r$--$L$ relation from Dietrich et al.\ (2009), i.e., Eq.~\ref{dietrichD.eq}.}} &
                         \multicolumn{1}{c}{$\frac{L_{bol}}{L_{edd}}$} &
                         \multicolumn{1}{c}{EW$_B$(\AA)} &
                         \multicolumn{1}{c}{EW$_{all}$(\AA)} \\
 \hline
   CSO 18	  & 1.300 &  139.87  &$ 2728.10  \pm 0.49 $&8.71      &9.25& 0.32 &20.28&$ 22.09\pm  0.37$ \\ 
   CSO 21	  & 1.190 &  151.63  &$ 3244.49  \pm 0.82 $&8.88      &9.41& 0.24 &22.80&$ 24.65\pm  0.33$ \\ 
   CSO 233	  & 2.030 &  220.37  &$ 1670.96  \pm 0.95 $&8.40      &8.92& 1.08 &17.60&$ 17.60\pm  0.84$ \\ 
   CSO 879	  & 0.549 &   44.02  &$ 1548.88  \pm 0.50 $&7.90(8.85)&8.50& 0.56 &10.47&$ 14.96\pm  0.33$ \\ 
   PG 0935$+$416  & 1.966 &  711.91  &$ 2518.34  \pm 1.19 $&9.08      &9.53& 0.86 & 4.68&$  5.00\pm  0.34$ \\ 
   PG 0946$+$301  & 1.220 &  355.37  &$ 2216.44  \pm 0.37 $&8.78      &9.27& 0.78 &13.62&$ 14.80\pm  0.27$ \\ 
   PG 1206$+$459  & 1.155 &  607.67  &$ 2441.60  \pm 0.76 $&9.01      &9.47& 0.84 &12.30&$ 13.46\pm  0.21$ \\ 
   PG 1248$+$401  & 1.032 &  254.23  &$ 2337.98  \pm 0.28 $&8.73      &9.24& 0.59 &16.82&$ 18.84\pm  0.30$ \\ 
   PG 1254$+$047  & 1.018 &  231.12  &$ 3603.36  \pm 0.57 $&9.08      &9.60& 0.24 &15.65&$ 18.63\pm  0.28$ \\ 
   PG 1259$+$593  & 0.472 &   76.12  &$ 2417.71  \pm 0.65 $&8.44(8.65)&9.01& 0.30 & 8.77&$ 12.71\pm  0.28$ \\ 
   PG 1338$+$416  & 1.204 &  249.35  &$ 3062.60  \pm 0.85 $&8.96      &9.47& 0.34 &13.22&$ 16.08\pm  0.29$ \\ 
   PG 1630$+$377  & 1.478 &  612.30  &$ 2793.42  \pm 1.91 $&9.13      &9.59& 0.65 &12.12&$ 13.45\pm  0.26$ \\ 
   Q 1628.5$+$3808& 1.461 &  193.09  &$ 2167.82  \pm 0.58 $&8.59      &9.12& 0.60 &16.54&$ 16.54\pm  0.37$ \\ 
   Q J0751$+$2919 & 0.912 &  347.62  &$ 2432.96  \pm 0.37 $&8.85      &9.34& 0.64 &14.79&$ 15.98\pm  0.25$ \\ 
   UM 497	  & 2.022 &  342.38  &$ 1901.77  \pm 0.56 $&8.63      &9.13& 1.04 &22.83&$ 22.83\pm  0.46$ \\ 
   US 1420	  & 1.473 &  199.32  &$ 1708.34  \pm 0.45 $&8.40      &8.92& 0.98 & 7.50&$ 10.76\pm  0.33$ \\ 
   US 1443	  & 1.564 &  302.09  &$ 2589.39  \pm 0.61 $&8.87      &9.37& 0.53 &14.10&$ 14.10\pm  0.30$ \\ 
   US 1498	  & 1.406 &  162.70  &$ 1844.98  \pm 0.29 $&8.41      &8.94& 0.76 &11.73&$ 12.87\pm  0.34$ \\ 
   US 1867	  & 0.513 &   39.58  &$ 1745.21  \pm 0.39 $&7.98(8.52)&8.58& 0.42 &13.12&$ 16.87\pm  0.24$ \\ 
   US 3150	  & 0.467 &   22.81  &$ 1467.23  \pm 1.09 $&7.68(8.55)&8.31& 0.45 & 8.84&$ 13.46\pm  0.39$ \\ 
   US 3472	  & 0.532 &   45.26  &$ 2234.99  \pm 0.39 $&8.23(8.94)&8.83& 0.27 &15.13&$ 18.94\pm  0.32$ \\ 
   TON 52	  & 0.434 &   28.10  &$ 1739.86  \pm 0.78 $&7.88(8.39)&8.51& 0.36 &17.31&$ 17.31\pm  0.49$ \\ 
   CSO 174	  & 0.653 &   36.73  &$ 3564.17  \pm 1.06 $&8.58(9.24)&9.19& 0.10 &32.41&$ 36.94\pm  0.36$ \\ 
   TON 34	  & 1.925 & 1761.85  &$ 3006.16  \pm 0.62 $&9.47      &9.88& 0.94 &10.77&$ 14.04\pm  0.28$ \\
   CSO 179	  & 0.782 &   95.95  &$ 2691.86  \pm 0.28 $&8.59(9.18)&9.15& 0.28 &20.44&$ 25.62\pm  0.33$ \\
   PG 1522$+$101  & 1.328 &  765.86  &$ 3437.96  \pm 0.90 $&9.37      &9.82& 0.48 &15.66&$ 16.13\pm  0.25$ \\
\hline
\end{tabular}
\end{minipage}
\end{table*}
\subsection{\mgii\ doublet region fit}
For fitting the \mgii doublet region, we consider a rest wavelength
range between 2200 {\bf and} 3200 \AA. The continuum in this region is
modeled by a power law, i.e., $a_{1} \lambda ^{-\alpha}$. For fitting
the \mgii\ doublet we have tried both Gaussians and truncated
5-parameter Gauss-Hermite series (Salviander et al.\ 2007; Dong et
al.\ 2009a).  In Gaussian profile fits, our initial guess consists of
two Gaussian components for each line of the \mgii\ doublet; however,
whenever the second component is not statistically required the
procedure automatically drops it during the fit.  Only two cases
where single components fits were found to be sufficient were detected (UM
479 and US 1443).  To test for any overfitting caused by
assuming two components, we also forced our procedure to fit only
single components; however, in doing so for all other sources, a very good
fit is found neither for the wings of the lines nor for their central
narrow cores. The truncated 5-parameter Gauss-Hermite series was also
tried, but the fits in this case also were not as good as those from
the two component Gaussian profile model, so our final analyses
employ that fitting approach.  \par

  In our two component Gaussian profile fits, the redshift and
width of each component (narrow/broad) of \mgiia\ were tied to the
respective components of the \mgiib line. In addition, we have constrained
the width of narrow component to be smaller than 1000 \kms\, and the
width of the broader \mgii component to be same as width of UV 
\feii\ emission line in the region (UV \feii). We used an UV \feii
template  generated by Tsuzuki et al.\ (2006), basically
from the measurements of I\,ZW\,1, which also employ  calculations
with the CLOUDY photoionization code (Ferland et al.\ 1998).  This
template is scaled and convolved to the FWHM value equivalent to the
broad components of \mgii\ by taking into account the FWHM of the
I\,ZW\,1 template. The best fit value of the broad component of
\mgii\ obtained in this way is finally used in our calculation of
BH mass.  The emission lines other than \feii lines identified
from the composite SDSS QSO spectrum (see Table 2 in Vanden Berk et
al.\ 2001), are modeled with single Gaussians. The final fit is
achieved by varying all the free parameters simultaneously by
minimizing the \chisq\ value, until the reduced $\chi^2_r$ is $\approx 1$.
Demonstrations of our spectral fitting in the UV region are given in
Fig.~\ref{lab:fig_mgiidemo}, and values for FWMHs and EWs are given
in Table~\ref{lab:tabmgii}.

\subsection{Optical microvariability and spectral properties}
 Among the 19 sources for which we have spectral
coverage of the \hbeta line, optical microvariability is shown by 7 of
them while the other 12 {\bf have not been seen to} show this property (Carini et al.\ 2007).  
Similarly, among the 26 sources with
spectral coverage of \mgii doublet, only 5 show optical
microvariability but the other 21  do not.\par

We have used our best fit values of FWHM and EW to search for any
correlation between them and optical microvariability.  The median
value of total rest frame EW of the \hbeta line (i.e., the sum of the
EWs of all fitted components) for 12 sources without optical
microvariability is found to be $\sim 69$\,\AA.  Its value for the 7
sources having optical microvariability is $\sim 79$\,\AA.  This seems
to be contrary to the expectation based on a blazar component model of
microvariability for RQQSOs, which would predict smaller EW values for
variable sources, due to dilution of emission line strength by jet
components (Czerny et al.\ 2008).  However, it should be noted that
observed \hbeta profiles certainly require multiple components to fit
them. Therefore, if optical microvariability is a phenomenon related
to the central AGN engine, then it is presumably better to do such a
comparison only using the broad component of the \hbeta line 
(Section~\ref{subsec:hbeta}).  This is because the clouds responsible
for this broad component are clearly in the sphere of influence of the
massive BH, while the other components may not be. The median value of
rest frame EW of this component for objects showing any optical
microvariability is $26.8$\,\AA\, and for objects without optical
microvariability it is $20.8$\,\AA, which is again contrary to the
expectation based on a blazar component model of optical
microvariability. The corresponding values of median FWHM of these broad
components for sources with and without microvariability are somewhat
different, at $4509$ km s$^{-1}$ and $4326$ km s$^{-1}$,
respectively.\par

 Similar to what we did with the \hbeta line, we have also used our
 \mgii spectral fits to look for any correlation with
 microvariability. The median value of rest frame EW of broad \mgii
 doublet components (with width tied to the UV \feii template width)
 is found to be $9.15$\,\AA\, for the five sources with optical
 microvariability and $9.24$\,\AA\, for the 21 sources without any
 confirmed microvariability.  These values are essentially identical,
 and so again do not support the simple blazar component model for
 microvariability in RQQSOs.  The corresponding values of median FWHM
 for sources with and without microvariability are $2432$ km s
 $^{-1}$ and $2447$ km s$^{-1}$, respectively, and so are
 indistinguishable.\par

 In Fig.~\ref{lab:histo_ewfwhm} we show the histograms of our FWHM and
 rest frame EW values based on our best fits for \hbeta and \mgii
 lines. The shaded and non-shaded regions correspond respectively to
 sources with and without confirmed optical microvariability. As the
 figure shows, the distributions of sources with and without
 microvariability are on the whole quite similar, {\bf though there
seems to be a hint that the median
 EW width of the \hbeta line might be larger for sources with optical
 microvariability compared to those without optical microvariability.
To quantify this posibility we performed Kolmogorov-Smirnov tests on
all the distributions shown in Fig.~\ref{lab:histo_ewfwhm}.
The probabilities that the  sources
with and without confirmed optical microvariability 
are drawn from similar distributions are: 0.613 for 
$\rm H\beta(\rm EW)$; 0.664 for
$\rm Mg~II(\rm EW)$; 0.792 for $\rm H\beta(\rm FWHM)$
and 0.995 for $\rm Mg~II(\rm FWHM)$. As these
null probabilities are all high our data samples
 do not establish or support any relation of FWHM or EW with the
 optical microvariability properties.  The lack of a difference in the EW values does not
support simple models in which the microvariability arises from jets.}
\section{Black Hole Mass Measurements}
\label{sec:bhmass}
 The strong M$_{BH}$-$\sigma$ correlation between supermassive black
 hole (SMBH) mass and the velocity dispersion of the surrounding
 spheroidal stellar system (e.g., Tremaine et al.\ 2002) have shown
 that black hole (BH) assembly is intimately related to the bulge properties of its
 host galaxy. This paradigm of ``BH--galaxy co-evolution'' boosted the
 efforts to measure BH masses over a wide range (e.g., Richstone
 et al.\ 1998; Nelson 2000; Bower et al.\ 2000). However, one cannot
 obtain BH mass estimates from either stellar or gas dynamics, based
 upon traditional methods, beyond $\sim$100 Mpc, due to the difficulty in
 resolving the BH sphere of influence.  For these more distant sources
 the only viable way to probe BH masses is through their AGN
 activity.\par

The reverberation mapping technique (e.g., Blandford \& McKee 1982;
Peterson 1993) is the most direct method to probe gas dynamics on
spatial scales close to the BH in AGN, but the required long-duration
spectroscopic monitoring campaigns currently preclude the method from
being applied to large numbers of objects.  Because the time-lags
involved grow linearly with the SMBH mass, only relatively nearby
Seyfert galaxies with modest values of $M_{\rm BH}$ have been studied
in this fashion.  Therefore for any large statistical studies one has
to rely on the more indirect technique of virial single-epoch methods,
which were pioneered by Dibai (1980) and shown to be consistent with
reverberation mapping masses (e.g., Bochkarev \& Gaskell 2009).  These
measurements make use of the widths of emission lines such as \hbeta,
\mgii or \civ in conjunction with the bolometric luminosity (e.g., Vestergaard
\& Peterson 2006).  This method relies upon the FWHM indicating the
typical velocities of clouds bound to the BH and the assumption that
the mean distances, $r$, of the clouds emitting particular lines are
determined by their ionization potentials, and thus scale with the
ionizing continuum, $L$, of the QSO.  We have used our \hbeta or \mgii
FWHM measurements to compute the BH mass of RQQSOs in our sample, as
such estimates will also impose important constraints in studies
trying to understand the nature of RQQSOs (e.g., Czerny et
al.\ 2008).\par

Vestergaard \& Peterson (2006) presented  improved empirical 
relationships useful for estimating the central BH mass based on 
FWHMs of emission lines such as \hbeta, \mgii and \civ. We used their
following scaling relation to estimate the BH 
masses based on \hbeta\ lines (Vestergaard et al.\ 2006, their Eq.\ 5)
  \begin{eqnarray}
  \lefteqn{\log \,M_{\rm BH} (\rm H\beta)  =
   \log \,\left[ \left(\frac{\rm FWHM(H\beta)}{1000~km~s^{-1}} \right)^2\right]} \nonumber \\
  & & \mbox{} + (6.91\pm0.02)+ \log \, \left( \frac{\lambda \it L_{\lambda} 
    {\rm (5100\,\AA)}}{10^{44} \rm erg~s^{-1}}\right)^{0.50\pm0.06}~, 
\label{logMopt_L51.eq}
  \end{eqnarray}
  where $L_{\lambda}{\rm (5100\,\AA)}$  is
  the monochromatic luminosity at 5100\,\AA, which we have computed
  from the best fit power-law continuum, $a_{1} \lambda ^{-\alpha}$, in our
  simultaneous fit of the whole spectral region (see Section~\ref{subsec:hbeta}). \par
 
    As is evident from this equation, careful measurements of the FWHM of
  emission lines are crucial for good BH mass estimations. So it is
  very important to decompose the lines into broad and narrow components, since
  the former are much more likely to be in the BH sphere of influence, while the later are
  contributed from the much more distant narrow line region under the dynamical
influence of the galaxy's stellar
  component. Therefore, as described in Section~\ref{subsec:hbeta}, we have
  constrained the narrow component's profile to be linked to the
\oiiiab line profile
  while the broad component's profile is associated with the broad \feii emission
  multiplet. We have modeled the whole \feii emission multiplet  by
  also decomposing it into broad and narrow component
  contributions. This aspect of the fitting has been largely ignored in
  almost all previous BH mass estimation studies until recent efforts by Dong et
  al.\ (2008, 2009a). So in the above Eq.~\ref{logMopt_L51.eq} the FWHM we used, from
  among our Gaussian de-composition of \hbeta\ line (Section~\ref{sec:aalysis}), is
  the one with its width tied with the width of the broad \feii\ emission line,
  after correcting for instrumental broadening. The result of our
  analysis using \hbeta lines is summarized in Table~\ref{lab:tabhbeta}.
\begin{figure}
\epsfig{figure=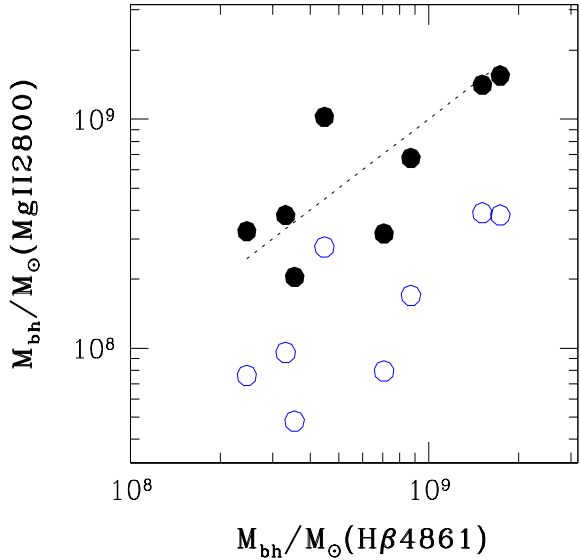,height=8.cm,width=8.cm,angle=00,bbllx=17bp,bblly=435bp,bburx=290bp,bbury=691bp,clip=true}
\caption[]{Comparison of BH masses estimated using \hbeta and \mgii lines for the eight 
  sources for which spectra cover both \hbeta and \mgii lines. The \mgii based
  BH masses using Eq.~\ref{logMuv_L30.eq} (McLure \& Dunlop 2004)
  are shown by hollow circles, while those based on Eq.~\ref{dietrichD.eq}
  (Dietrich et al.\ 2009) are shown by filled circles.  The dotted line shows 
  equal  BH masses estimated using  \hbeta and \mgii lines.
}
\label{lab:fig_compare}
\end{figure}
   
  To estimate black hole masses based on \mgii, we initially  adopted the
calibrations in McLure \& Dunlop (2004), i.e.,
 \begin{eqnarray}
    \lefteqn{\log \,M_{\rm BH} (\rm Mg~II) = (0.62\pm 0.14) \log \left( \frac{\lambda \it L_{\lambda}
        \rm (3000\,\AA)} {10^{44} \rm erg~s^{-1}} \right) }\nonumber \\
    & & \mbox{} + 2 \log \left(\frac{\rm FWHM(Mg~II)}{\rm km~s^{-1}} \right) + 0.505 ~,
    \label{logMuv_L30.eq}
  \end{eqnarray}
where the monochromatic luminosity at 3000\,\AA\, has been computed from
our power-law fit of the continuum and the FWHM used is the broad
component of our double Gaussian decomposition model, after correcting
for the instrumental broadening. The result of our \mgii\ analysis for
black hole mass based on the McLure \& Dunlop (2004) scaling relation is
summarized in Column 5 of Table~\ref{lab:tabmgii}.\par

 In our sample eight QSO spectra cover both \mgii and \hbeta lines,
 which allow us to compare whether or not the black hole masses
 estimated with them are consistent with one another. As can be noted
 from column 5 of Table~\ref{lab:tabmgii}, where BH mass estimates
 based on \hbeta lines are in parentheses, the differences can be even
 as large as 0.95 dex (e.g., CSO 879,
 Fig.~\ref{lab:fig_hbetademo2}). Also, we note that in all eight cases the
 \mgii based mass estimate is smaller than the one based on the \hbeta
 line.  One possible reason could be the uncertainty in the slope,
 $\beta$ of the $r$--$L$ relation; from Eq.\ \ref{logMuv_L30.eq}, $\beta
 =0.62\pm0.14$, so the nominal uncertainty is $\sim 25$ per cent.  Recent
 studies on the $r$--$L$ relation also indicate that
 the slope of the radius--luminosity relation is close to $\beta =
 0.5$, consistent with basic predictions of photoionization models
 (e.g., Davidson \& Netzer 1979; Osterbrock \& Ferland 2005). The
 presence of steeper slopes in some analyses appear to be caused by
 not correcting for contributions of the host galaxy to the continuum
 luminosity (Bentz et al.\ 2006; Bentz et al.\ 2009). Dietrich et
 al.\ (2009) have addressed this problem by assuming the slope of the
 $r$--$L$ relation to be 0.5, as seems to hold for the \hbeta line,
 and they then calculated the average scaling constant $D$; following
 their Eq.\ (6), viz,

\begin{eqnarray}
 M_{bh}(H\beta ) &=& M_{bh}(\rm Mg~II) = D\,
     \biggl({\lambda L_\lambda (3000) \over {10^{46} {\rm erg\,s}^{-1}}}\biggr)^{0.5}\, \nonumber \\
       & & \biggl({FWHM({\rm MgII}\lambda 2800) \over {1000\, {\rm km\,s}^{-1}}}\biggr)^2
     M_\odot\ ~,
\label{dietrichD.eq}
\end{eqnarray}
where the M$_{bh}$(H$\beta$) values were computed from their
near-infrared spectra and the FWHM(MgII$\lambda$ 2800) values came
from their optical spectra. They find $D = (2.0\pm0.5)\times
10^8\,M_\odot$.  Using this empirical constant $D$ in
Eq. ~\ref{dietrichD.eq}, we recalculated the \mgii -based BH masses;
the results are given in the sixth column of Table~\ref{lab:tabmgii}.  As can be noted from columns (5) and (6) and illustrated in
Fig.\,\ref{lab:fig_compare}, the discrepancy between the masses from the
two lines for the cases where both can be evaluated is reduced, and
ranges from 0.05 to 0.36 dex, which is within the expected systematic
uncertainty of black hole masses (around a factor of 4) based on
single epoch measurements (Vestergaard \& Peterson 2006).  These
consistent BH mass estimates range from $\sim 10^8$ to $>10^9~M_\odot$
for our sample.  We note that the prescriptions we have adopted are
not unique and that several other prescriptions for black hole masses
in terms of virial parameters obtained from \halpha as well as \hbeta
and \mgii also have been considered.  These have been recently
summarized by McGill et al.\ (2008) who show that the level of
agreement we have obtained can be improved upon somewhat with the
consideration of a larger number of AGN.
\begin{figure*}
\epsfig{figure=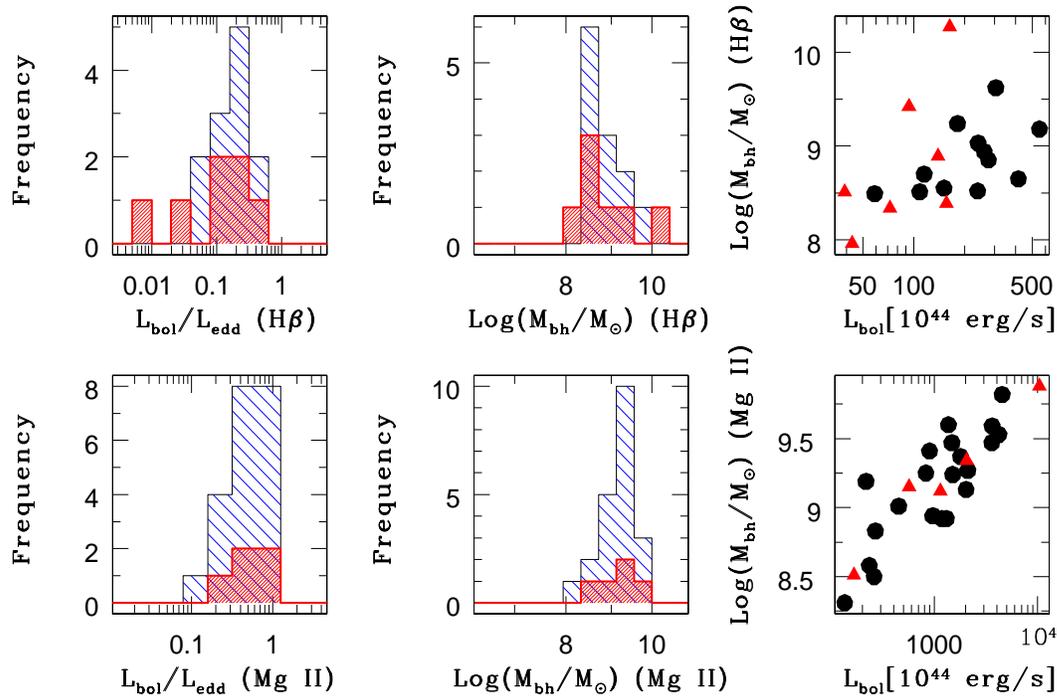,height=10.cm,width=15.cm,angle=00,bbllx=10bp,bblly=338bp,bburx=590bp,bbury=716bp,clip=true}
\caption[]{The upper left and middle panels show the histograms of
Eddington ratio $L_{bol}/L_{edd}$ and ${\rm log}(M_{bh}/M_{\odot})$
based on the \hbeta line, while the corresponding bottom panels show the results
based on the \mgii line. The shaded regions correspond to sources with
confirmed optical microvariability while the non-shaded regions are for
those sources for which optical microvariability is not detected.
The right upper and bottom panels respectively show plots of ${\rm log}(M_{bh}/M_{\odot})$
versus $L_{bol}$.
Triangles and circles respectively show
sources with and without confirmed optical microvariability.}
\label{lab:fig_eddi_ratio}
\end{figure*}
\begin{figure*}
\epsfig{figure=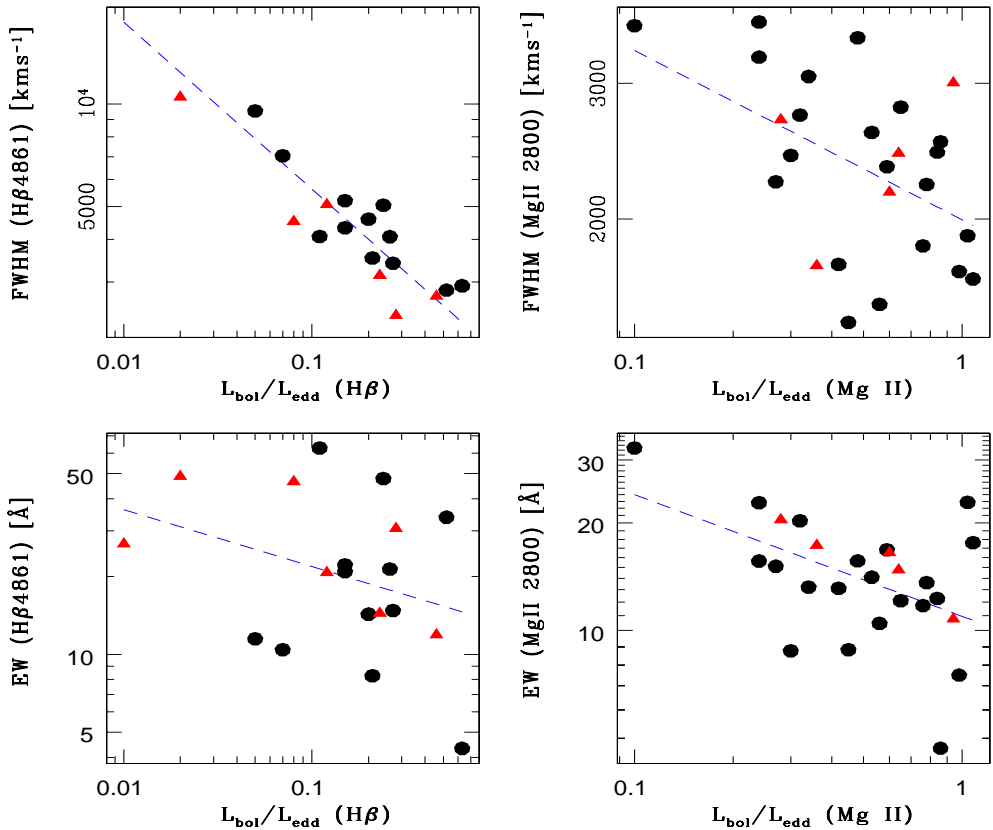,height=12.cm,width=14.cm,angle=0}
\caption[]{Observed variation of FWHM (upper) and EW (lower) with
  Eddington ratio ($\ell={\rm L}_{bol}/{\rm L}_{edd}$) based on both
  \hbeta (left) and \mgii (right) lines.  Triangles and circles
  respectively show sources with and without confirmed optical
  microvariability. {\bf The dashed lines show the linear regression fits
  treating $\ell$ as the independent variable. The
   correlations of both FWHM and EW with $\ell$,  
  suggest that the Eddington ratio is one of the fundamental parameters
  responsible for some AGN properties}.}
\label{lab:lratio_ew_fwhm}
\end{figure*}

\begin{figure*}
\epsfig{figure=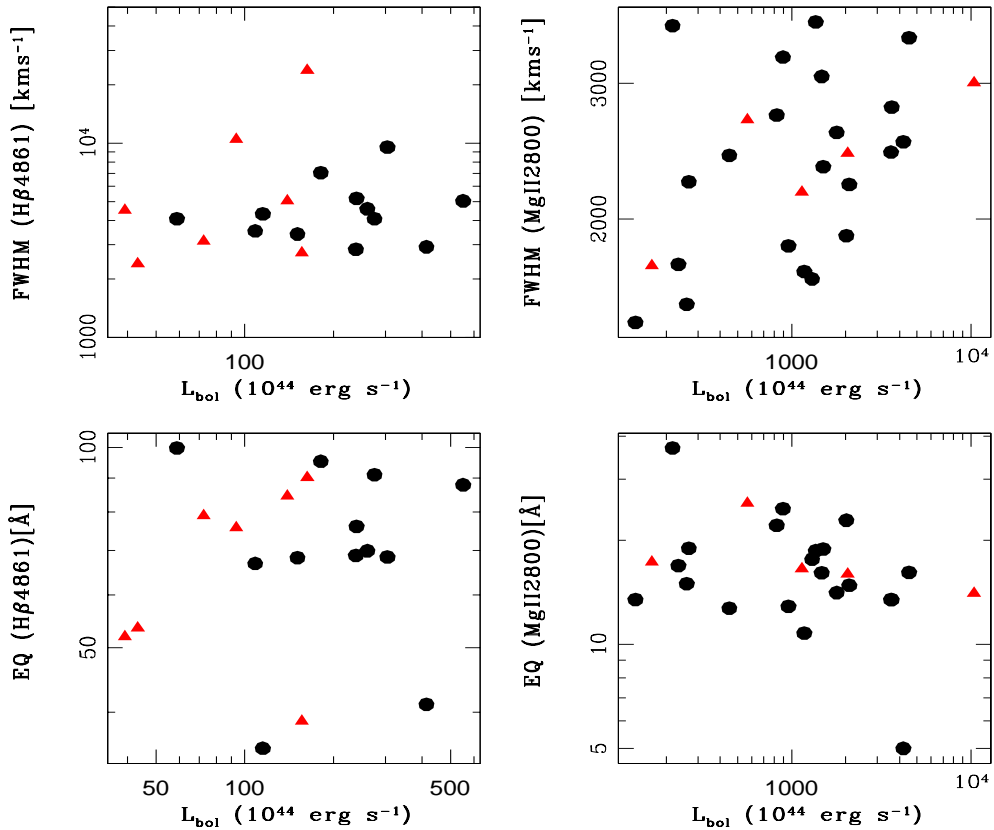,height=12.cm,width=14.cm,angle=0}
\caption[]{As in Fig.~\ref{lab:lratio_ew_fwhm} with $L_{\rm bol}$ as the
independent variable. This panel  indicates that most of the sources with optical microvariability
are of low luminosities.}
\label{lab:lbolo_ew_fwhm}
\end{figure*}
\begin{table*}
 \centering
 \begin{minipage}{140mm}
  \caption{BH growth time estimates (in units of Gyr)  from the H$\beta$ line analysis}
 \label{lab:grhbeta}

  \begin{tabular}{@{}lc|ccc|ccc|c@{}}

\hline

\multicolumn{1}{c}{object}&
\multicolumn{1}{c}{z}&
\multicolumn{3}{c|}{$L_{bol}$/$L_{edd}$\,=\,1.0}&
\multicolumn{3}{c|}{$L_{bol}$/$L_{edd}$\,obs.\footnote{Values of $\ell$ estimated
                       using H$\beta$ lines (Table~\ref{lab:tabhbeta}).}}&
\multicolumn{1}{c}{age of the}\\
\multicolumn{1}{c}{ }&
\multicolumn{1}{c}{ }&
\multicolumn{3}{c}{$M_{bh}$(seed)}&
\multicolumn{3}{c}{$M_{bh}$(seed)}&
\multicolumn{1}{c}{Universe}\\
\multicolumn{1}{c}{ }&
\multicolumn{1}{c}{ }&
\multicolumn{1}{c}{$10\,M_\odot$}  &
\multicolumn{1}{c}{$10^3\,M_\odot$}&
\multicolumn{1}{c}{$10^5\,M_\odot$}&
\multicolumn{1}{c}{$10\,M_\odot$}  &
\multicolumn{1}{c}{$10^3\,M_\odot$}&
\multicolumn{1}{c}{$10^5\,M_\odot$}&
\multicolumn{1}{c}{{[}$10^9$\,yr]} \\
\hline
CSO 174	        &0.653 &0.83 & 0.63 & 0.43 & 11.81  &  8.94 &  6.07 &  7.44  \\
 CSO 179	&0.782 &0.82 & 0.62 & 0.42 &  3.42  &  2.58 &  1.75 &  6.73  \\
CSO 448	        &0.317 &0.74 & 0.54 & 0.33 &  3.20  &  2.33 &  1.46 &  9.89  \\ 
CSO 879	        &0.551 &0.79 & 0.59 & 0.39 &  3.03  &  2.26 &  1.49 &  8.08  \\ 
Mrk 1014  	&0.163 &0.70 & 0.50 & 0.30 &  2.49  &  1.78 &  1.06 & 11.43  \\ 
PG 0832$+$251	&0.331 &0.79 & 0.59 & 0.39 &  6.59  &  4.92 &  3.25 &  9.77  \\ 
PG 0923$+$201	&0.192 &0.84 & 0.64 & 0.44 & 42.22  & 32.19 & 22.16 & 11.12  \\
PG 0931$+$437	&0.458 &0.86 & 0.66 & 0.46 & 17.29  & 13.28 &  9.27 &  8.73  \\ 
PG 1049$-$005   &0.359 &0.81 & 0.60 & 0.40 &  5.37  &  4.03 &  2.69 &  9.52  \\ 
PG 1259$+$593 	&0.474 &0.77 & 0.57 & 0.37 &  1.22  &  0.90 &  0.58 &  8.62  \\ 
PG 1307$+$085	&0.154 &0.75 & 0.55 & 0.35 &  6.83  &  5.01 &  3.18 & 11.53  \\
PG 1309$+$355	&0.183 &0.77 & 0.57 & 0.37 &  5.15  &  3.81 &  2.47 & 11.22  \\
PG 1444$+$407 	&0.268 &0.75 & 0.55 & 0.35 &  3.59  &  2.63 &  1.68 & 10.35  \\ 
Q 1252$+$0200 	&0.344 &0.93 & 0.73 & 0.53 & 92.97  & 72.91 & 52.85 &  9.66  \\ 
TON 52	  	&0.434 &0.74 & 0.54 & 0.34 &  1.61  &  1.18 &  0.74 &  8.91  \\
US 1867	        &0.514 &0.75 & 0.55 & 0.35 &  1.45  &  1.06 &  0.68 &  8.33  \\ 
US 3150	        &0.469 &0.76 & 0.56 & 0.36 &  2.80  &  2.06 &  1.32 &  8.65  \\
US 3472	        &0.532 &0.80 & 0.60 & 0.40 &  3.98  &  2.98 &  1.98 &  8.20  \\
US 995     	&0.227 &0.75 & 0.55 & 0.35 &  9.41  &  6.91 &  4.40 & 10.76  \\

\hline
\end{tabular}
\end{minipage}
\end{table*}
\begin{table*}
 \centering
 \begin{minipage}{140mm}
  \caption{BH growth time estimates (in units of Gyr) from the \mgii line analysis}
   \label{lab:grmgii}

  \begin{tabular}{@{}lc|ccc|ccc|c@{}}

\hline
\multicolumn{1}{c}{object}&
\multicolumn{1}{c}{z}&
\multicolumn{3}{c|}{$L_{bol}$/$L_{edd}$\,=\,1.0}&
\multicolumn{3}{c|}{$L_{bol}$/$L_{edd}$\,obs.\footnote{Values of $\ell$ 
    estimated using  \mgii lines (Table~\ref{lab:tabmgii}).} }&
\multicolumn{1}{c}{age of the}\\
\multicolumn{1}{c}{ }&
\multicolumn{1}{c}{ }&
\multicolumn{3}{c}{$M_{bh}$(seed)}&
\multicolumn{3}{c}{$M_{bh}$(seed)}&
\multicolumn{1}{c}{Universe}\\
\multicolumn{1}{c}{ }&
\multicolumn{1}{c}{ }&
\multicolumn{1}{c}{$10\,M_\odot$}  &
\multicolumn{1}{c}{$10^3\,M_\odot$}&
\multicolumn{1}{c}{$10^5\,M_\odot$}&
\multicolumn{1}{c}{$10\,M_\odot$}  &
\multicolumn{1}{c}{$10^3\,M_\odot$}&
\multicolumn{1}{c}{$10^5\,M_\odot$}&
\multicolumn{1}{c}{{[}$10^9$\,yr]} \\
 \hline
    CSO 18	   & 1.300 & 0.83&  0.63&  0.43& 2.59 & 1.96 & 1.33 & 4.73\\ 
    CSO 21	   & 1.190 & 0.84&  0.64&  0.44& 3.51 & 2.68 & 1.84 & 5.07\\ 
    CSO 233	   & 2.030 & 0.79&  0.59&  0.39& 0.74 & 0.55 & 0.36 & 3.18\\ 
    CSO 879	   & 0.549 & 0.75&  0.55&  0.35& 1.34 & 0.98 & 0.63 & 8.09\\ 
    PG 0935$+$416  & 1.966 & 0.86&  0.65&  0.45& 0.99 & 0.76 & 0.53 & 3.28\\ 
    PG 0946$+$301  & 1.220 & 0.83&  0.63&  0.43& 1.06 & 0.81 & 0.55 & 4.97\\ 
    PG 1206$+$459  & 1.155 & 0.85&  0.65&  0.45& 1.01 & 0.77 & 0.53 & 5.18\\ 
    PG 1248$+$401  & 1.032 & 0.83&  0.63&  0.43& 1.40 & 1.06 & 0.72 & 5.63\\ 
    PG 1254$+$047  & 1.018 & 0.86&  0.66&  0.46& 3.59 & 2.76 & 1.92 & 5.68\\ 
    PG 1259$+$593  & 0.472 & 0.80&  0.60&  0.40& 2.68 & 2.01 & 1.34 & 8.63\\ 
    PG 1338$+$416  & 1.204 & 0.85&  0.65&  0.45& 2.50 & 1.91 & 1.32 & 5.02\\
    PG 1630$+$377  & 1.478 & 0.86&  0.66&  0.46& 1.33 & 1.02 & 0.71 & 4.25\\
    Q 1628.5$+$3808& 1.461 & 0.81&  0.61&  0.41& 1.36 & 1.02 & 0.69 & 4.30\\
    Q J0751$+$2919 & 0.912 & 0.84&  0.64&  0.44& 1.31 & 0.99 & 0.68 & 6.12\\
    UM 497	   & 2.022 & 0.82&  0.61&  0.41& 0.78 & 0.59 & 0.40 & 3.19\\
    US 1420	   & 1.473 & 0.79&  0.59&  0.39& 0.81 & 0.61 & 0.40 & 4.27\\
    US 1443	   & 1.564 & 0.84&  0.64&  0.44& 1.58 & 1.21 & 0.83 & 4.05\\
    US 1498	   & 1.406 & 0.80&  0.60&  0.40& 1.05 & 0.78 & 0.52 & 4.44\\
    US 1867	   & 0.513 & 0.76&  0.56&  0.36& 1.81 & 1.33 & 0.85 & 8.33\\
    US 3150	   & 0.467 & 0.73&  0.53&  0.33& 1.63 & 1.18 & 0.74 & 8.67\\
    US 3472	   & 0.532 & 0.79&  0.58&  0.38& 2.91 & 2.17 & 1.42 & 8.20\\
    TON 52	   & 0.434 & 0.75&  0.55&  0.35& 2.09 & 1.53 & 0.98 & 8.92\\
    CSO 174	   & 0.653 & 0.82&  0.62&  0.42& 8.21 & 6.21 & 4.20 & 7.43\\
    TON 34	   & 1.925 & 0.89&  0.69&  0.49& 0.95 & 0.73 & 0.52 & 3.35\\
    CSO 179	   & 0.782 & 0.82&  0.62&  0.42& 2.92 & 2.20 & 1.49 & 6.73\\
    PG 1522$+$101  & 1.328 & 0.88&  0.68&  0.48& 1.84 & 1.42 & 1.01 & 4.65\\
\hline
\end{tabular}
\end{minipage}
\end{table*}
\section{Eddington ratios and black hole growth times}
We have also estimated the Eddington ratio $\ell \equiv L_{bol}/L_{edd}$, where
$L_{bol}$ is taken as $  5.9 \times \lambda L_{\lambda}$(3000\AA)  and 
$9.8 \times \lambda L_{\lambda}$(5100\AA)  for \mgii and \hbeta,
respectively (McLure \& Dunlop 2004), and $L_{edd}=1.45 \times 10^{38}
(M_{bh}/M_{\odot}) \rm erg~s^{-1} $, assuming a mixture of hydrogen
and helium so the mean molecular weight is $\mu=1.15$. 
The results are given in Fig.\,\ref{lab:fig_eddi_ratio} which shows that distributions
of sources showing and not showing optical microvariability are
not significantly different with respect to either BH mass or $\ell$. 

To test the reasonableness of estimated  Eddington ratios, we also 
computed black hole growth times to compare them with the age of the 
Universe (at the time the QSO is observed), by using the following equation 
(Dietrich et al.\ 2009),
\begin{equation}
  M_{bh}(t_{obs}) = M_{bh}^{seed}(t_0) \,
   {\rm exp}\biggl(\ell \,{(1-\epsilon) \over \epsilon}\,{\tau \over  t_{edd}}\biggr)~,
\label{growthtau.eq}
\end{equation}
\noindent
where $\tau = t_{obs} - t_0$ is the time elapsed since the initial
time, $t_0$, to the observed time, $t_{obs}$; $M_{bh}^{seed}$ is the
seed BH mass; $\epsilon$ is the efficiency of converting mass to
energy in the accretion flow, and $t_{edd}$ is the Eddington time scale, with $t_{edd}
=\sigma_T c / 4 \pi G m_p = 3.92 \times 10^8$\,yr (Rees 1984). We used
Eq.~\ref{growthtau.eq} to derive the times, $\tau$, necessary to
accumulate the BH masses listed in Tables~\ref{lab:tabhbeta} and
\ref{lab:tabmgii}, for seed black holes with masses of $M_{bh}^{seed}
= 10 M_\odot$, $10^3 M_\odot$, and $10^5 M_\odot$, respectively. Two
cases are considered: (i) BHs are accreting at the Eddington-limit,
i.e., $\ell $\,=\,L$_{bol}$/L$_{edd}$\,=\,1.0 and the efficiency of
converting mass into energy is $\epsilon=0.1$; (ii) BHs are accreting
with our observed Eddington ratios and $\epsilon=0.1$.  These results
are summarized in Tables~\ref{lab:grhbeta} and ~\ref{lab:grmgii}.

{\bf In Fig.~\ref{lab:lratio_ew_fwhm} we show the observed variations of FWHM
and EW with $\ell$ based on both \hbeta and \mgii lines. There
appear to be linear relations of both FWHM and EW with $\ell$ in these log-log plots. 
For the FWHM plots this is unsurprising since the BH masses are proportional 
to $L_{edd}$. To quantify any such linear relations for the EW plots we
perform linear regressions, treating $\ell$ as the independent variable, and find
\begin{eqnarray}
 \rm logEW(H\beta)=(1.12\pm0.14)+(-0.22\pm0.15)\rm log \ell , \nonumber \\
 \rm logEW(\rm Mg~II) =(1.04\pm0.05)+(-0.34\pm0.12)\rm log \ell .
\label{regr_hbeta.eq}
\end{eqnarray}
 Here the errors on the fit parameters are purely statistical.
We have calculated the 
Spearman rank correlations of
logEW with log$\ell$, and found the correlation coefficient for 
r$_{s}(H\beta)=-0.22$,  with null probability $p_{null}=0.35$, and so only a possible
weak correlation is present.  Whereas r$_{s}(\rm Mg~II)=-0.52$, 
with $p_{null}=0.01$ and so this negative correlation is significant.}

\section{Discussion and Conclusions}

Several recent papers have used the very large numbers of quasars
discovered by modern surveys to estimate their BH masses using virial
approaches (e.g., Shen et al.\ 2008; Fine et al.\ 2008; Vestergaard \&
Osmer 2009).  The large numbers of quasars whose spectra are fit in
these particular papers (between 1100 and almost 57,700) have allowed 
for very important new conclusions to be drawn about quasar demography.  Shen
et al.\ (2008) have found that the line widths of \hbeta and \mgii
follow lognormal distributions with very weak dependencies on redshift
and luminosity.  Their comparison of BH masses of radio-loud quasars
(RLQSOs) and BALs with those of ``ordinary" quasars (RQQSOs) shows that
the mean of virial masses of RLQSOs is 0.12 dex larger than that of
ordinary quasars, while that of BALs is indistinguishable from that of
ordinary quasars. Fine et al.\ (2008) used \mgii lines to estimate BH
masses, and found that the scatter in measured BH masses is
luminosity dependent, showing less scatter for more luminous
objects. Vestergaard \& Osmer (2009) have studied the mass
functions of BH masses at different redshifts, and found evidence for
cosmic downsizing in their cosmic space density distributions.

The sample we have considered is much smaller, but each member has
been carefully selected to be among the special group of RQQSOs
and Seyfert galaxies already examined for optical microvariability
(e.g., Carini et al.\ 2007).  This criterion demands that modest
aperture (usually 1--2 m) telescopes can make precise photometric
measurements in just a few minutes, and so limits the members to the
rare QSOs with bright apparent magnitudes (usually $m_V < 17.5$).  In
addition, special care was taken in the fitting of the line profiles
as discussed in Sections 3.1 and 3.2.

As the precision of CCD based differential photometry has typically
improved to better than 0.01 mag for these  measurements made in a few minutes,
the question as to the presence of microvariability in RQQSOs has been
clearly settled in the affirmative (Gopal-Krishna et al.\ 2000,
2003; Stalin et al.\ 2004a,b; Gupta \& Joshi 2005; Carini et al.\ 2007; 
Gupta \& Yuan 2009; Ram{\'i}rez et
al.\ 2009).  However these papers find that the duty cycle (DC), or
fraction of nights such behavior is detected, is low, roughly 10--20\%,
for RQQSOs observed for reasonably long periods (4 hours or more)
during a night.  On the other hand, blazars show much more rapid
variability, with DCs for such monitoring periods around 50--60\% and when
observed for $>6$ hours the blazar DCs rise to 60--85\% (e.g., Carini
1990; Sagar et al.\ 2004; Gupta \& Joshi 2005).  The difference in microvariabilty
amplitudes and DCs between RQQSOs and blazars can be easily understood
if all the fluctuations arise from a jet very close to the central
engine, but that jet only escapes that region and emits significant radio
power for RL objects (e.g.,
Gopal-Krishna et al.\ 2003; Gopal-Krishna, Mangalam \& Wiita 2008).  
And most RLQSOs that are not
blazars would only have modest (if any) Doppler boosting, so 
that they also have DCs comparable to those of RQQSOs, as found in
the samples of Stalin et
al.\ (2005) and Ram{\'i}rez et al.\ (2009), or at a  level between those
of RQQSOs and blazars (35--40\%, in the sample of Gupta \& Joshi 2005),
is not surprising.  Only recently has a decent sized sample of radio intermediate quasars (RIQs),
defined as those relatively rare objects with radio to optical flux
ratios in the regime between those that are truly radio quiet and
those normally defined as RLQSOs, been targeted for microvariability
monitoring (Goyal et al.\ 2009); their key result is that the RIQs also
probably have a DC $\leq$ 20\%.  Therefore they are not likely to be
beamed versions of RQQSOs as has been frequently suggested (Goyal et
al.\ 2009).

We conclude that there {\bf may be a} weak negative correlation between \hbeta
EW and the Eddington ratio, $\ell$, but there is a {\bf significant} one
between the \mgii EW and $\ell$ {\bf (Fig.~\ref{lab:lratio_ew_fwhm}; Section~5)}.
  This latter point has 
been made independently and more firmly by Dong et al.\ (2009b) using
a larger sample drawn from SDSS. We can also see from Fig.\ 7 that
there is a decline in FWHM with $\ell$; this is unsurprising since the BH
masses are proportional to $L_{edd}$.  We see from Fig.\ 8 that there
is a tendency for sources with detectable optical microvariability to
have somewhat lower luminosities than those with no such detections.
This is interesting, and should be confirmed by examining larger
samples.  But it is not surprising, as regardless of whether the
microvariability arises in accretion discs or jets, more massive BHs
have naturally longer characteristic timescales that will presumably
make any variations harder to detect over the course of a single
night.

  We also find that the BH masses
estimated from the FWHMs of both the \hbeta and \mgii lines are
reasonable, in that growth to their estimated masses from even small
seed BHs is easily possible within the age of the Universe at their
observed redshift if the mean $\ell$ values are close to unity (Tables
4 and 5).  This remains true for the great majority of RQQSOs if the
value of $\ell$ we compute from the current continuum flux was
constant until the time we observe them; however, this assumption does
not work for 4 out of the 19 QSOs with \hbeta lines.  It is also
problematical for 1 out of the 26 QSOs with \mgii lines (i.e., CSO 174,
which is also among the difficult set of 4 using \hbeta profiles).

We find no difference between the EWs (of both the \hbeta or \mgii
lines) and the presence or absence of radio emission in those QSOs
(Figs.\ 4 and 7).  As discussed in the introduction, if much of the
optical emission in RLQSOs comes from a jet, then we would expect the
EWs of the RLQSO sources to be significantly lower than those of the
RQQSOs, and if anything, they are slightly higher in our sample.  This
result does not support the hypothesis (e.g., Czerny et al.\ 2008)
that RQQSOs possess jets that are producing rapid variations.  Instead
it may indicate that variations involving the accretion disc (e.g., Wiita 2006) play an important role here.

Improvements to our results could be obtained through extensive searches for INOV in a
larger sample of RQQSOs.  It should be useful to divide the
sample of sources  based on their EWs as available from SDSS, or otherwise uniformly obtained, spectra.
Such samples could be divided into three classes based on EW of \hbeta,
e.g.,  EW$<$40\AA, 40\AA $\le$ EW $<$ 80\AA\, and EW $\ge$
80\AA. Such samples should be made as homogeneous as possible on the basis of apparent
magnitudes, $z$ and M$_{V}$. With such larger and homogeneous samples,
the absence of a correlation between EW and DC of INOV, as found here in our modest
sample, could be confirmed or shown to be unlikely.

\section*{Acknowledgments}

PJW is grateful for hospitality at ARIES; this work was supported in part by a subcontract to GSU from NSF grant AST05-07529 to the University of
Washington.

  Funding for the SDSS and SDSS-II has been provided by
the Alfred P. Sloan Foundation, the Participating Institutions, the
National Science Foundation, the U.S. Department of Energy, the
National Aeronautics and Space Administration, the Japanese
Monbukagakusho, the Max Planck Society, and the Higher Education
Funding Council for England. The SDSS Web Site is
http://www.sdss.org/.

    The SDSS is managed by the Astrophysical Research Consortium for
the Participating Institutions. The Participating Institutions are
the American Museum of Natural History, Astrophysical Institute
Potsdam, University of Basel, University of Cambridge, Case
Western Reserve University, University of Chicago, Drexel
University, Fermilab, the Institute for Advanced Study, the Japan
Participation Group, Johns Hopkins University, the Joint Institute
for Nuclear Astrophysics, the Kavli Institute for Particle
Astrophysics and Cosmology, the Korean Scientist Group, the
Chinese Academy of Sciences (LAMOST), Los Alamos National
Laboratory, the Max-Planck-Institute for Astronomy (MPIA), the
Max-Planck-Institute for Astrophysics (MPA), New Mexico State
University, Ohio State University, University of Pittsburgh,
University of Portsmouth, Princeton University, the United States
Naval Observatory, and the University of Washington.

\label{lastpage}
\end{document}